\documentclass{article}

\usepackage[nonatbib,final]{neurips_2019}

\usepackage[utf8]{inputenc} 
\usepackage[T1]{fontenc}    
\usepackage{hyperref}       
\usepackage{url}            
\usepackage{booktabs}       
\usepackage{amsfonts}       
\usepackage{nicefrac}       
\usepackage{microtype}      
\usepackage{graphicx}
\usepackage{caption}
\usepackage{tabularx}
\usepackage{float}
\usepackage{subfig}
\usepackage[english]{babel}

\usepackage[numbers,square,sort&compress]{natbib}
\title{Modeling Product Search Relevance in e-Commerce}

\author{%
  Rahul Radhakrishnan Iyer\thanks{Corresponding Author: \texttt{rahuli@alumni.cmu.edu}} \\
  Language Technologies Institute\\
  Carnegie Mellon University\\
  Pittsburgh, PA 15213 \\
  \texttt{rahuli@andrew.cmu.edu} \\
   \And
  Rohan Kohli \\
  Language Technologies Institute\\
  Carnegie Mellon University\\
  Pittsburgh, PA 15213 \\
  \texttt{rkohli1@andrew.cmu.edu} \\
   \AND
  Shrimai Prabhumoye \\
  Language Technologies Institute\\
  Carnegie Mellon University\\
  Pittsburgh, PA 15213 \\
  \texttt{sprabhum@andrew.cmu.edu} \\
}

\begin{document}

\maketitle

\begin{abstract}
With the rapid growth of e-Commerce, online product search has emerged as a popular and effective paradigm for customers to find desired products and engage in online shopping. However, there is still a big gap between the products that customers really desire to purchase and relevance of products that are suggested in response to a query from the customer. In this paper, we propose a robust way of predicting relevance scores given a search query and a product, using techniques involving machine learning, natural language processing and information retrieval. We compare conventional information retrieval models such as BM25 and Indri with deep learning models such as word2vec, sentence2vec and paragraph2vec. We share some of our insights and findings from our experiments.
\end{abstract}

\section{Introduction}
\label{sec:intro}
E-Commerce has seen an unprecedented growth in recent years, resulting in a vast number of customers engaging in online shopping to find desired products. With such an increase in adoption, it becomes pertinent for online retailers to be able to suggest products that are very relevant to a customer's search query. We propose a model to be able to predict search relevance scores given a search query and a product. Our dataset which is described in the following section, is  taken from Kaggle, and focuses on search relevance prediction, specific to an e-Commerce organization, Home Depot. The input to our system will be pairs of product IDs and search queries; the output will be search relevance score, on a scale of $1$-$3$, for each pair. The evaluation criteria for the performance of our model is going to be the root mean-squared error (RMSE) and the correlation coefficient of the predicted relevance scores. Since we do not have multiple products for each query, our focus is therefore on the relevance alone and not on the rankings, which is why we do not perform a ranking evaluation.

\subsection{Related Works}
\label{subsec:related}
There is plenty of published literature about techniques for free-text querying for information retrieval. Most of the popular techniques that have been proposed for search relevance have been aimed at web search. Some of the older works on information retrieval that are relevant to our goal have revolved around exploring different ways for modeling language (vector space models, query likelihood models, variations of popular systems like Lucene, Okapi BM25, etc.) \citep{perez2009integrating}

A few popular methods (such as Indri \citep{strohman2005indri}) have been inspired by Bayesian networks. While many of these information retrieval models are inspired by very different underlying theory, most of them in practice utilize a few common features such as using tf-idf \citep{ramos2003using}, bag of words approaches / unigram features, treating documents and queries as vectors and using similarity metrics such as cosine similarity \citep{tan2006mining} and KL divergence, using simple maximum likelihood estimation approaches as well as maximum a-priori estimates using query-independent evidence, etc. These features have over the last few years become a central part of information retrieval. There have been other works on structured contextual clustering \citep{roustant2005structured} and optimizing web search using web click-through data \citep{xue2004optimizing}. Recently, several approaches involving natural language processing \cite{iyer2019event,iyer2019unsupervised,iyer2019heterogeneous,iyer2017detecting,iyer2019machine,iyer2017recomob,iyer2019simultaneous}, machine learning \cite{li2016joint,iyer2016content,honke2018photorealistic,iyer2020correspondence}, deep learning \cite{iyer2018transparency,li2018object} and numerical optimizations \cite{radhakrishnan2016multiple,iyer2012optimal,qian2014parallel,gupta2016analysis,radhakrishnan2018new} have also been used in the visual and language domains.

The motivation and goal behind our paper is to make better use of machine learning for improving product-search relevance. Compared to existing works, we are much more focused on exploiting the small amount of product data that is available to us (consisting solely of the natural language product details) by trying out new representations of the data and new learning techniques to automatically learn these representations. The evaluation criteria here is also different since we are only interested in predicting absolute relevance scores (which are more important in the context of an e-commerce environment) rather than the direct ranking of results. 

Recent developments in the field of deep learning have popularized the use the of word embeddings for various NLP tasks which aim to capture the contextual meaning of natural language. We aim to use such methods in predicting product relevance, see how they fare in comparison to more conventional approaches, and whether they can be combined in a meaningful way.

\section{Data Set}
\label{sec:data}

The training data is available to us in a relational form and is divided into the following tables:
\begin{enumerate}
\item Training data : This includes the product uid (primary key / number), product title, search query and the relevance score (which is also the label. This is a floating point number between 1.0 and 3.0). This table consists of $74067$ instances.
\item Test data: This consists of $166693$ instances.
\item Product description: This consists of the product uid(foreign key to the training data table) and the product description.
\item Product attributes: This consists of product uid (foreign key), attribute name and the attribute value. 
\end{enumerate}

Each product that is a part of the train and test data has exactly one corresponding description. A product may have any number of attributes (or none at all). Most products have different sets of attributes that are specific to the product being considered.

\subsection{Collection}
\label{subsec:collection} 
The process of data collection was quite straight-forward since the complete data for training was available to us through the \textit{kaggle} website in the form of comma separated values. 

\subsection{Preprocessing}
\label{subsec:preprocess}
We concatenated various text features from different tables to create a matrix of feature vectors and performed some basic pre-processing on the textual features: case-conversion to lowercase, stop-word removal (using a list of stop-words from NLTK) and stemming (using a porter stemmer) and tokenization (creating n-gram features out of the remaining tokens). Furthermore, on the basis of error analysis on initial experiments, we also employed a basic spelling correction algorithm based on the Levenshtein distance between new unseen tokens and dictionary terms. We also converted numerical characters and number into a canonical form by a simple regular expression mechanism.

\begin{table}
\vspace{2ex}
\begin{tabular}{l | c}
\toprule
\textbf{Measure} & \textbf{Value}\\
\midrule
Number of unique products & 65535\\
Number of unique unigram words in the product description data & 50274\\
Range for count of words in product description data & [5 to 71093]\\
Number of unique unigram words in the product title data & 9644\\
Range for count of words in product title data & [5 to 41573]\\
Number of unique unigram words in the search term data & 6158\\
Range for count of words in search term data & [5 to 2652]\\
\bottomrule
\end{tabular}
\centering
\caption{Some statistics of the corpus}
\label{table:stats}
\end{table}

\subsection{Corpus Exploration}
We have performed an in-depth analysis of the corpus and some of the key statistics computed have been summarized in  Table \ref{table:stats}

\section{Methodology}
\label{sec:method}

\subsection{Baselines}
\subsubsection{N-gram models and Boolean Retrieval}
For our baseline experiments we used two methods. First we used a unigram model i.e. extracted unigram features for the search term, product title and product description. Secondly, we computed OR and AND scores of the search term with the product title and the product description. These operators are explained in detail below. On both these methods of extracting feature sets, we used the Sequential Minimal Optimization (SMO) \citep{platt1998sequential} Regression algorithm for training the SVM. The SMO is an effective way for solving the quadratic programming (QP) problem that arises during the training of support vector machines and is very commonly used in information retrieval systems. We chose to work with RBF (Radial Basis Function) kernel for the SVM. The different parameter values that were chosen for the method are listed in Table \ref{table:svm_params}.
We have  used 10-fold cross validation and generated relevance scores for the test data between 1.0 to 3.0. We chose to work with SVM because:
\begin{enumerate}
\item We get to choose from multiple Kernels
\item We can engineer the kernel (in our case find the optimal Gamma value) so that the model gains more knowledge about the features.
\item SVM has a regularization parameter through which we can control over-fitting.
\item SMO in particular is very successful in solving the convex optimization problem by which the SVM is defined.
\end{enumerate}
Authors in \citep{van2004benchmarking} have thoroughly benchmarked the SVM classifiers.

\begin{table}
\vspace{2ex}
\begin{tabular}{l | c | l}
\toprule
\textbf{Parameter} & \textbf{Value} & \textbf{Explanation}\\
\midrule
c & 1.0 & Complexity Parameter\\
$\gamma$ & 0.01 & Gamma value in the RBF kernel\\
$\epsilon$ & 0.001 & Epsilon parameter of the epsilon insensitive loss function (Regression)\\
tolerance & 0.001 & Stopping criteria for convergence\\
\bottomrule
\end{tabular}
\centering
\caption{Parameters used for the SVM}
\label{table:svm_params}
\end{table}

We have tuned the $\gamma$ parameter in the RBF kernel. We considered values $\in \lbrace 10^{-10}, 10^{-9}, \ldots, 1 \rbrace$, and we found that the best performing parameter value is 0.01. 

\subsubsection{OR, AND and NEAR operators}
If we have a search term of n words ($q_1$, $q_2$, $\ldots$, $q_n$) and document $d_j$ then we calculate the OR score $q_{OR}$($q_1$, $q_2$, $\ldots$, $q_n$) on $d_j$ as:
\\
score($q_{OR}$($q_1$, $q_2$, $\ldots$, $q_n$), $d_j$) =  MAX(score($q_1$, $d_j$), ($q_2$, $d_j$), $\ldots$, score($q_n$, $d_j$))
\\
We calculate the count of each word in the search term in the product title, product attributes and product description individually. We take the maximum count of the words in search term. eg: If my search query is `angle bracket' and the product title is `Simpson Strong-Tie 12-Gauge Angle' then the OR score is 1.
\\
score($q_{AND}$($q_1$, $q_2$, $\ldots$, $q_n$), $d_j$) =  MIN(score($q_1$, $d_j$), ($q_2$, $d_j$), $\ldots$, score($q_n$, $d_j$))
\\

\subsection{Baseline results}
We tried various representations of data as part of baseline experiments. For n-gram models (with the Markov assumption), we extracted unigram, bigram and trigram features from the data. For the learning algorithm, we used SVM to train a regression model (Model 1). SVM and n-gram models gave reasonably good performance with an RMSE in the range of  0.2880 to 0.3057.  However, since we found that bigram and trigram models were quite computationally expensive to train on the entire training data, we performed most experiments on a small subset of the data using just unigrams. This is the reason we have not reported the results for bigrams/trigrams. We found that increasing the amount of training data from 10,000 to 74067 instances resulted in a marginal improvement, but at a very high computational cost. 

In contrast to the n-gram features, we find that features inspired by common information retrieval models such as Boolean retrieval (OR, AND) to be much more useful in improving the model. We again trained a regression model (Model II) using just 6 of these features (described earlier) alone and got a much better RMSE of 0.2872.

Finally, we tried to took the predictions from Model 1 and use those as an additional feature for Model 2 to train another model (Model III). This actually resulted in a statistically significant improvement. We found the third model to be better than any of the other models. It gives a very good RMSE performance, takes into account the most useful n-gram features (by using the relevance predictions of Model 1) and at the same time has a very small feature space;  so is quite fast to train.

\subsection{More advanced information retrieval models : Okapi BM25 and Indri}
\begin{figure}
\captionsetup{width=0.8\textwidth}
  \centering
  \includegraphics[scale=0.5]{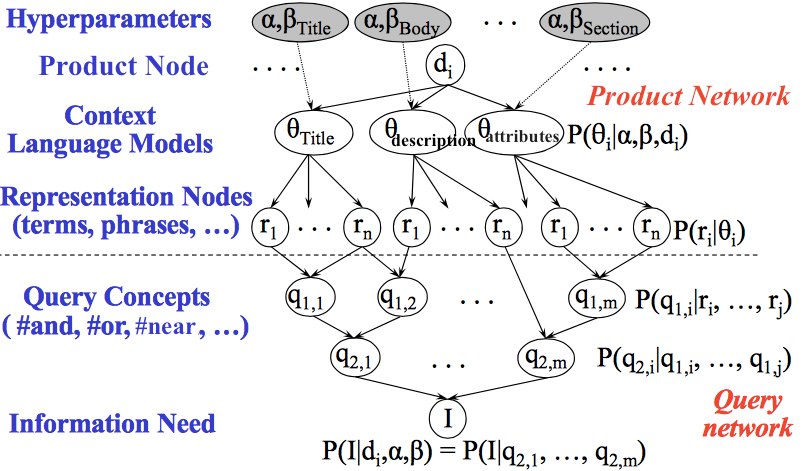}
  \caption{The figure shows the Bayesian network that the Indri model is based on. We create 3 language models for different kinds of product data, and get the frequency counts for each of them (representation nodes). The first layer of the query network ($q_1$, x) shows the raw scores calculated for each query term. The second layer($q_2$, x) shows how these scores are combined.}
  \label{fig:indri}
\end{figure}

We calculate the BM25 and Indri scores for each instance. Indri and BM25 are two popular ranking functions based on a probabilistic retrieval framework and a Bayesian Network respectively \citep{robertson1995okapi}. 
The Indri and BM25 scores assign a relevance score to a <query, product> pair by computing a function of the query term frequency in a product description,  the ratio of the product's description length to the average product description length, and the number of products that the term appears in (tf-idf). The idea is that the relative length of the product description (that contains some matching keyword) acts as a prior since shorter descriptions with keyword matches are more likely to be more relevant to our query. The product frequency (no. of products that a query term is matched with) plays a simiar role. The Indri model is shown in the Figure \ref{fig:indri}.

We used two levels of smoothing in our calculation of Indri scores to avoid zero entries for the query terms: linear interpolation as the first level and Bayesian smoothing with Dirichlet priors as the second. Assuming $q$ is the search query in consideration and $q = \lbrace q_1, q_2, \ldots, q_n \rbrace$ are the $n$ search terms in the query, the smoothing used is given below:  
\begin{equation}
	p(q|d) = \prod\limits_{q_i \in q} \left( (1-\lambda) \frac{tf_{q_i, d} + \mu \frac{ctf(q_i)}{length(c)}}{length(d) + \mu} + \lambda \frac{ctf(q_i)}{length(c)} \right)
\end{equation}

where $d$ is the document in consideration (the product in our case), $tf_{q_i, d}$ is the term-frequency of the search term $q_i$ in document $d$, $ctf(q_i)$, the collection term frequency of $q_i$, is the frequency of $q_i$ across all the documents, $c$ is the length of the collection of all the documents, $\lambda$ is the parameter in the linear interpolation model and $\mu$ is the parameter in the Bayesian smoothing.

\subsection{Word and Sentence Embeddings}
One major problem in our baseline model which was evident from some cursory error analysis was that our baseline system relied a little too heavily on exact matched terms as being the strongest features for most instances. It failed to take into account the context of words. In order to overcome this problem, we decided to use word and sentence embeddings to represent our data – namely the textual features of product title,descriptions and attributes. These embeddings capture the context of the search term and product descriptions. 

Since most of the entries in our dataset have single words as search terms, we expect word vectors to be much more useful than sentence vectors. Nevertheless, we also decided to do experiments using sentence vectors which we felt could be helpful for products with lengthy descriptions and search terms. We have used the sentence to vector model proposed in \citep{kiros2015skip}. The Sentence to Vector model uses Recurrent Neural Network using Gated Recurrent Units for training their model. For our experiments we have used the word to vector model described in \citep{le2014distributed}. It uses the skipgram model to generate word embeddings.  We trained the google word to vector model on our dataset to generate vectors which capture the context of our dataset.

\section{Experiments}
\label{sec:experiments}
\subsection{Evaluation Metrics}
In our experiments, we have used two different metrics to evaluate our performance: the root mean-squared error (RMSE) and the correlation coefficient of the predicted and gold standard relevance scores.

\subsection{Baseline Experiments}
We have performed experiments with varying amount of training samples. For each experiment, we use SMO model to predict scores and calculate the correlation and RMSE values using 10-fold cross validation. These are listed below in Table \ref{table:features}:

\begin{table}
\vspace{2ex}
\begin{tabular}{c | c | c | l}
\toprule
\textbf{\#Training Examples} & \textbf{\#Features Extracted} & \textbf{\#Features Used} & \textbf{Features}\\
\midrule
2000 & 5516 & 200 & Unigram\\
2000 & 6 & 6 & OR + AND\\
5000 & 9640 & 200 & Unigram\\
5000 & 6 & 6 & OR + AND\\
1000 & 14946 & 200 & Unigram\\
10000 & 6 & 6 & OR + AND\\
74067 & 6 & 6 & OR and AND\\
\bottomrule
\end{tabular}
\centering
\caption{Different Features Extracted: Feature Space Design}
\label{table:features}
\end{table}

The entire dataset of 74067 samples has 54042 unigram features.

\subsection{Sentence to Vector Model}
We first generate the embedding of the product description and the search term using the skip-thought vector model \citep{kiros2015skip}. We then train a logistic regression with sigmoid activation unit. The input feature for the model is the dot product of the embeddings of the product description and the search term. The Y labels for the model for training phase were the relevance score. During the testing phase, we get the features using the dot product and then we generate the Y label using the model. This is done by taking the dot product of the prediction probability of the class (In our case, we have 3 classes since the score is between 1-3) and the no of classes.

\subsection{Word to vector Model}
\begin{figure*}[!t]
\centering
\subfloat[Flowchart for the Word2Vec model]{\includegraphics[scale=0.5]{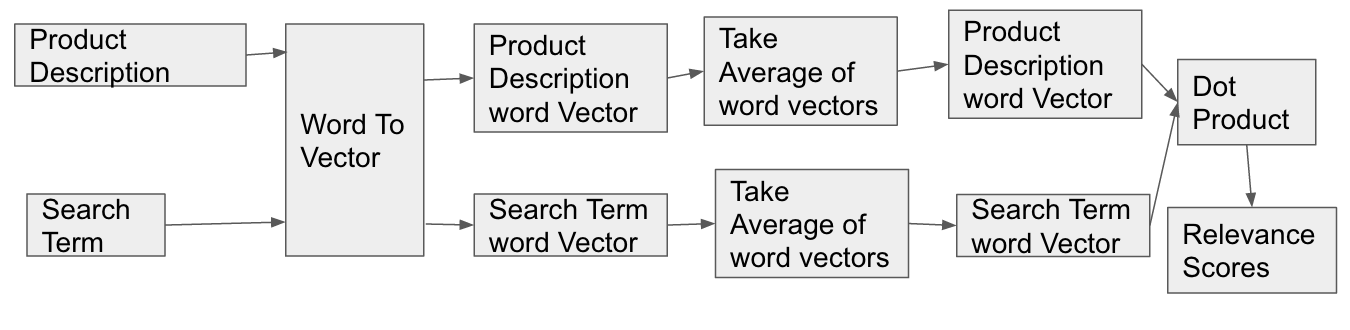}
\label{fig:sentence2vec}}
\hfil
\subfloat[Flowchart for the Sentence2Vec model]{\includegraphics[scale=0.5]{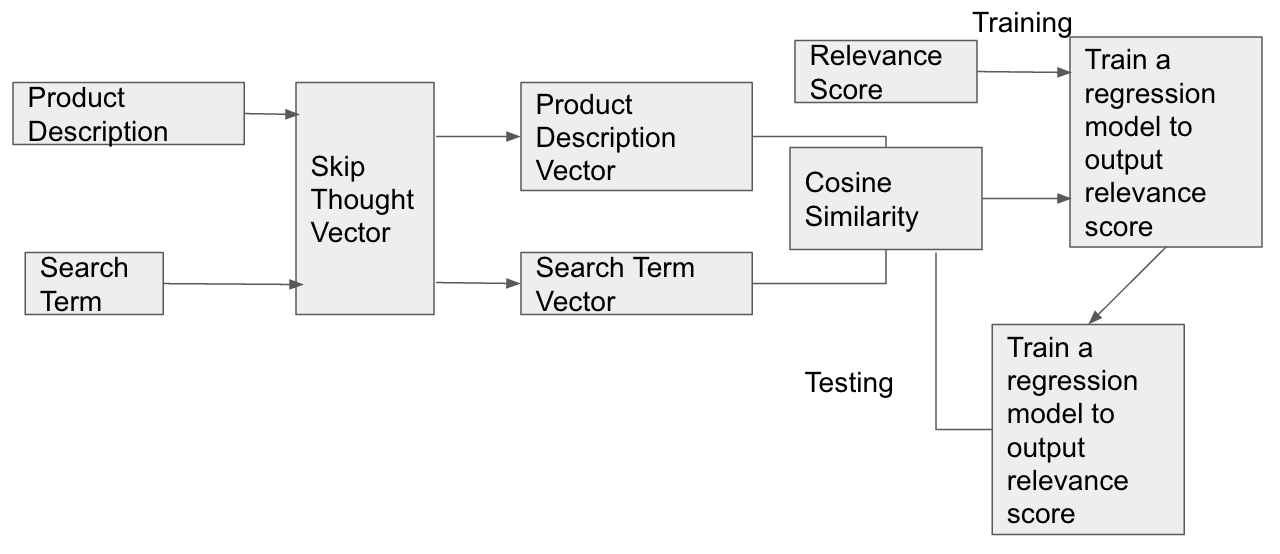}
\label{fig:word2vec}}
\caption{Flowcharts for the models Used}
\label{fig:flowcharts}
\end{figure*}

We have trained the Google word to vector model \citep{mikolov2013efficient} on our dataset. We generate the word embeddings for each word in the product description. We then take an average of the word vectors for it to capture the information of the product in one vector. We calculate the word embedding of each term of the search term and then average them as well. Then, we calculate the dot product between the averaged vector of product description and the search term to yield a score between -1 and 1. This is then made to fit in the range of 1 to 3, and is outputted as the relevance score. Figure \ref{fig:flowcharts} shows the architecture and flow for our experiments.

\subsection{Paragraph to vector}
We have used the gensim library\footnote{https://radimrehurek.com/gensim/}, which takes a document of text as input and generates its embedded vector. This was motivated from the model described in \citep{le2014distributed}. The product descriptions and the search terms were fed to it and in the process the embedded vectors were obtained. Once again we get the cosine similarity between this vector and the search term vector to predict the relevance score. 

\section{Results and discussion}
\label{sec:results}

\begin{figure*}[!t]
\centering
\subfloat[Comparison of Correlation Coefficient]{\includegraphics[scale=0.32]{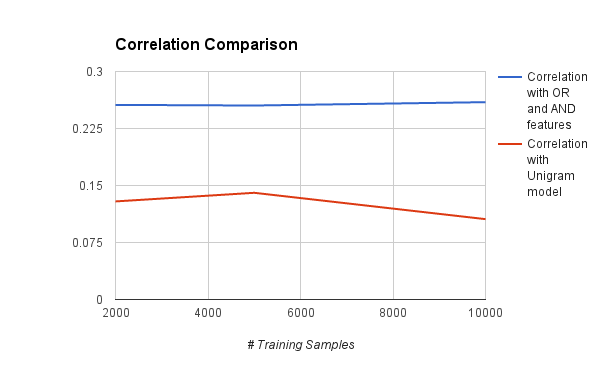}
\label{fig:correlation}}
\hfil
\subfloat[Comparison of RMSE value]{\includegraphics[scale=0.32]{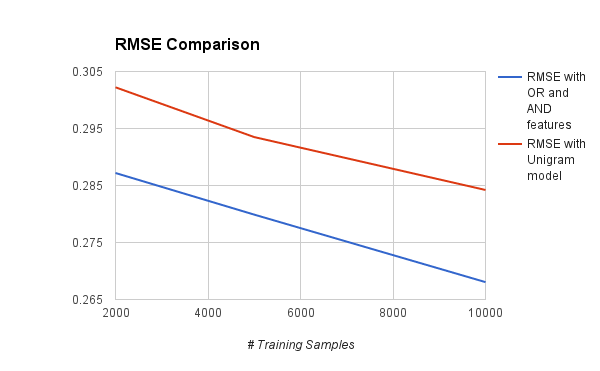}
\label{fig:rmse}}
\hfil
\subfloat[Comparison of Correlation Coefficient Across Different Models]{\includegraphics[scale=0.32]{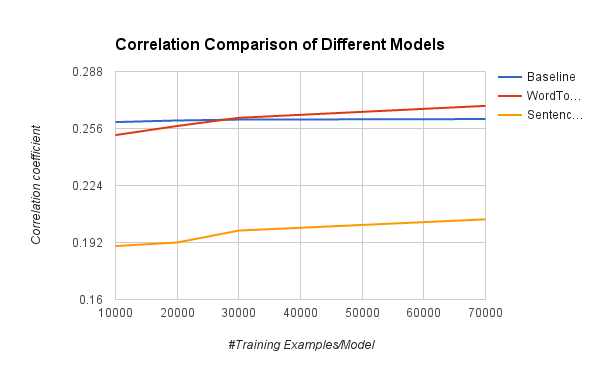}
\label{fig:corr_comp_models}}
\caption{Comparison of Correlation Coefficient and RMSE}
\label{fig:comparison_rmse_corr}
\end{figure*}

Figure \ref{fig:correlation} shows the comparison of the correlation coefficients between the Unigram type of features and the OR and AND features. As we can see, the correlation coefficient is much higher (better) using only the 4 features obtained from the OR and AND operator for all of the experiments. Figure \ref{fig:rmse} shows the comparison of the RMSE value between the Unigram type of features and the OR and AND features. Here, too we see that the RMSE value is lower (better) for the 4 features extracted from the OR and AND operator for all of the experiments. Figure \ref{fig:corr_comp_models} shows the comparison of correlation coefficients across the different models considered, where we see that there is a slight improvement in performance of Word2Vec over our baselines. The result values are shown in the Table \ref{table:results}. From the results, we see that the features calculated using BM25 and Indri models are much better in representing our data than the methods used in the baseline.

\begin{table}
\vspace{2ex}
\begin{tabular}{c | l | c | c}
\toprule
\textbf{\#Training Examples} & \textbf{Features} & \textbf{Correlation Coefficient} & \textbf{RMSE Value}\\
\midrule
2000 & Unigram & 0.1290 & 0.3022\\
2000 & OR + AND & 0.2560 & 0.2872\\
5000 & Unigram & 0.1404 & 0.2935\\
5000 & OR + AND & 0.2552 & 0.2799\\
1000 & Unigram & 0.1057 & 0.2842\\
10000 & OR + AND & 0.2596 & 0.2680\\
74067 & OR + AND & 0.2613 & 0.2657\\
\bottomrule
\end{tabular}
\centering
\caption{Baseline results obtained for various features}
\label{table:results}
\end{table}

In Table \ref{table:model_results} we have shown the RMSE and correlation coefficient value results for the deep network models and BM25 and Indri model. The table also shows the dimension of the vector / the number of features used by each model. We have calculated the Pearson correlation coefficient and RMSE value as described in [14,15]. During the training phase of the word to vector model we had a vocabulary size of 226705.

\begin{table}[ht]
\vspace{2ex}
\resizebox{\textwidth}{!}{\begin{tabular}{c | m{5cm} | l | l | c | c }
\toprule
\textbf{Example Id.} & \textbf{Product Description} & \textbf{Search Term} & \textbf{Model} & \textbf{Actual Relevance} & \textbf{Predicted Relevance}\\
\midrule
& & & & & \\
1. & This 1/2 in. x 260 in. PTFE Tape is compliant with federal specification T-27730A. The tape provides a strong tight seal for threaded joints and easily disassembles. Compliant with federal specification \ldots & ptfe & Sentence to vec & 2.67 & 1.991 \\
& & & & & \\\hline
& & & & & \\
2. & Enhance any doorway with this elegant 4 ft. Sparkling Pine Potted Artificial Christmas Tree, pre-lit with 70 clear incandescent lights for added shimmer.Measures 4 ft. tall with 22 in. base diameter \ldots & 4 foot christmas tree colored  lights & Sentence to vec & 2.33 & 2.39 \\
& & & & & \\\hline
& & & & & \\
3. & his 1/2 in. x 260 in. PTFE Tape is compliant with federal specification T-27730A. The tape provides a strong tight seal for threaded joints and easily disassembles. Compliant with federal specification t-27730a.Virgin white PTFE constructionUse on plastic, brass, copper, aluminum, galvanized-steel and black-iron piping (piping sold separately)Provides a strong tight seal for threaded joints \ldots & ptfe & Word to vec & 2.67 & 2.60  \\
& & & & & \\\hline
& & & & & \\
4. & his 1/2 in. x 260 in. PTFE Tape is compliant with federal specification T-27730A. The tape provides a strong tight seal for threaded joints and easily disassembles. Compliant with federal specification t-27730a.Virgin white PTFE construction Use on plastic, brass, copper, aluminum, galvanized-steel and black-iron piping (piping sold separately)Provides a strong tight seal for threaded joints \ldots & ptfe & Baseline & 2.67 & 1.882  \\
& & & & & \\\hline
& & & & & \\
5. & Enhance any doorway with this elegant 4 ft. Sparkling Pine Potted Artificial Christmas Tree, pre-lit with 70 clear incandescent lights for added shimmer.Measures 4 ft. tall with 22 in. base diameter \ldots & 4 foot christmas tree colored  lights & Word to vec & 2.33 & 1.99 \\
& & & & & \\
\bottomrule
\end{tabular}}
\centering
\caption{Error Analysis - some typical examples of correctly and incorrectly classified examples.}
\label{table:model_analysis}
\end{table}

Furthermore, we tried to combine the results of our best models using two different approaches (i.e. BM25/Indri/scores and the word2vec model similarity score) to train another SVM model. However, this did not result in any statistically significant improvement.

\begin{table}
\vspace{2ex}
\resizebox{\textwidth}{!}{\begin{tabular}{l | c | c | c}
\toprule
\textbf{Model} & \textbf{Correlation Coefficient} & \textbf{RMSE Value} & \textbf{Dimension of Vector} \\
\midrule
OR, AND, BM25, Indri & 0.2537 & 0.2509 & 6\\
Word To Vector & 0.2687  & 0.2482 & 100\\
Sentence To Vector & 0.205 & 0.2748 & 2400\\
Paragraph To Vector& 0.2302& 0.2706 & 100\\
SVM with Word2Vec Scores as Features & 0.2701 & 0.2496 & 7\\
\bottomrule
\end{tabular}}
\centering
\caption{Results obtained for various models}
\label{table:model_results}
\end{table}

In Table \ref{table:model_analysis}, we have shown how different models fail or succeed in different cases. The example 1. shows that when the search term is just a word like `ptfe' which does not mean anything by itself the sentence to vector model will not be able to capture the context of this word at all. Hence, the predicted score is much lower than the actual relevance score. Whereas in example 2.,  the search term is ` 4 foot christmas tree colored  lights'. Hence, the model already has some context about lawn mowers and covers. Hence, it captures the context and predicts a score very close to the actual relevance score. In example 3. we have applied word to vec model on example 1. and this model works better on this example because during training phase of the model, it captured information of `ptfe'. In example 4. we get baseline score for the example 3. The baseline performs worse than the word to vec model here because it just count the number of times `ptfe' occurs in the product description. Hence, the count is OR and AND score is 2. Hence it predicts a value lower than the actual relevance score. In example 5.  we have applied word to vec model on example 2. Here sentence to vec model performs better because the query term is long and in the form of a sentence and the sentence embedding captures the context of the query.

\section{Conclusion}
\label{sec:conclusion}

Through the baselines experiments we observed that simply using an SVM with features from different information retrieval models helps improve on the relevance scores predicted by some of the models alone. To address the problem of the context of words being ignored by the baseline model, we tried various methods that use deep networks. We found that the the word2vec model resulted in a significant improvement with an RMSE of 0.2482. In most cases, it did a good job in capturing the relevance of ambiguous queries correctly. On the other hand, we found that the sentence2vec / skip-thought vectors actually caused a slight decline in performance. It appears that the most likely reason for this is that many of the descriptions do not have very well-formed English sentences and the text is in more of a bullet point format consisting of many unrelated product characteristics written together. So this explains why using sentence vectors for such relevance tasks may not be a good idea. In contrast, word vectors still do a good job here.

However, we could not achieve any significant improvement when we tried to train a single SVM model using features from both our best models (i.e. BM25/Indri scores as well as similarity scores from the word2vec model). It is hard to explain this result but it's most probable that there are some spurious correlations among the features learned by these two approaches which confuses the SVM. 

Furthermore, since these models are computationally expensive to train, we compared the performance of our deep learning models trained on different sizes of training sets in order to get an estimate of how many training example might be sufficient to get a good enough performance. We found that the performance improves considerably while increasing the training data and our results indicate that we can expect a much more drastic improvement if we are able to get hold of more labeled data in order to train a better model.

\newpage

\bibliography{sample}

\end{document}